\documentclass[a4paper,fleqn,usenatbib,useAMS]{mnras}
\usepackage{color}
\usepackage{graphicx}
\usepackage{pdflscape}	
\usepackage{amsmath}	
\usepackage{amssymb}	
\usepackage{multicol}        
\usepackage{bm}		
\usepackage{ulem}
\usepackage{ascii}

\usepackage{times}

\newcommand{\Ha}{\ifmmode \text{H}\alpha \else H$\alpha$\fi\xspace}
\newcommand{\Hb}{\ifmmode \text{H}\beta \else H$\beta$\fi\xspace}
\newcommand{\oi}{\ifmmode [\text{O}\,\textsc{i}] \else [O~{\scshape i}]\fi\xspace}
\newcommand{\Oi}{\ifmmode [\text{O}\,\textsc{i}]\lambda 6300 \else [O~{\scshape i}]$\lambda 6300$\fi\xspace}
\newcommand{\oii}{\ifmmode [\text{O}\,\textsc{ii}] \else [O~{\scshape ii}]\fi\xspace}
\newcommand{\Oii}{\ifmmode [\text{O}\,\textsc{ii}]\lambda 3726 + \lambda 3729 \else [O~{\scshape ii}]$\lambda 3726 + \lambda 3729$\fi\xspace}
\newcommand{\nii}{\ifmmode [\text{N}\,\textsc{ii}] \else [N~{\scshape ii}]\fi\xspace}
\newcommand{\Nii}{\ifmmode [\text{N}\,\textsc{ii}]\lambda 6584 \else [N~{\scshape ii}]$\lambda 6584$\fi\xspace}
\newcommand{\oiii}{\ifmmode [\text{O}\,\textsc{iii}] \else [O~{\scshape iii}]\fi\xspace}
\newcommand{\Oiii}{\ifmmode [\text{O}\,\textsc{iii}]\lambda 5007 \else [O~{\scshape iii}]$\lambda 5007$\fi\xspace}
\newcommand{\sii}{\ifmmode [\text{S}\,\textsc{ii}] \else [S~{\scshape ii}]\fi\xspace}
\newcommand{\Sii}{\ifmmode [\text{S}\,\textsc{ii}]\lambda 6716 + \lambda 6731 \else [S~{\scshape ii}]$\lambda 6716 + \lambda 6731$\fi\xspace}
\newcommand{\wha}{\ifmmode W_{\mathrm{H}\alpha} \else $W_{\mathrm{H\alpha}}$\fi\xspace}
\newcommand{\ww}{\ifmmode W2 - W3 \else $W2 - W3$\fi\xspace}

\defcitealias{CidFernandes2010}{CF10}
\defcitealias{CidFernandes2011}{CF11}
\defcitealias{Kewley2001}{K01}
\defcitealias{Kauffmann2003c}{K03}
\defcitealias{Stasinska2006}{S06}

\usepackage[T1]{fontenc}
\usepackage{ae,aecompl}

\title[The many faces of LINER-like galaxies: a WISE view]
      {The many faces of LINER-like galaxies: a WISE view}

\author[Herpich et al.]
{ F.\ Herpich$^{1}$\thanks{E-mail:herpich@astro.ufsc.br}, 
  A.\ Mateus$^{1}$, 
  G.\ Stasi\'nska$^{2}$,
  R.\ Cid Fernandes$^{1}$,
  N.\ Vale Asari$^{1}$
  \\
  $^{1}$Departamento de F\'{\i}sica--CFM, Universidade Federal de Santa Catarina, C.P.\ 476, 88040-900, Florian\'opolis, SC, Brazil \\
  $^{2}$LUTH, Observatoire de Paris, PSL, CNRS, UMPC, Univ Paris Diderot, 5 place Jules Janssen, 92195 Meudon, France
}

\date{Accepted 2016 July 15. Received 2016 July 14; in original form 2016 March 09}

\pubyear{2016}

\begin{document}
\label{firstpage}
\pagerange{\pageref{firstpage}--\pageref{lastpage}}
\maketitle

\begin{abstract}
We use the SDSS and WISE surveys to investigate the real nature of  galaxies defined as LINERs in the BPT diagram. After establishing a mid-infrared colour $\ww = 2.5$ as the optimal separator between galaxies with and without star formation, we investigate the loci of different galaxy classes in the  \wha versus \ww  space. We find that: (1) A large fraction of LINER-like galaxies are emission-line retired galaxies, i.e galaxies which have stopped forming stars and are powered by  hot low-mass evolved stars (HOLMES). Their \ww colours show no sign of star formation  and their \Ha\ equivalent widths, \wha, 
are consistent with ionization by their old stellar populations. 
(2) Another important fraction have \ww indicative of star formation. This includes objects located in the  supposedly `pure AGN' zone of the BPT diagram. (3) A smaller fraction of LINER-like galaxies have no trace of star formation from \ww and a high \wha, pointing to the presence of an AGN. (4) Finally, a few LINERs tagged as retired by their \wha but with \ww values indicative of star formation are late-type galaxies whose SDSS spectra cover only the old `retired' bulge. This reinforces the view that LINER-like galaxies are a mixed bag of objects involving different physical phenomena and observational effects thrusted into the same locus of the BPT diagram. 
\end{abstract}


\section{Introduction}

Since 1981, inferring the main excitation sources of galaxies generally relies on diagrams involving emission line ratios \citep*{Baldwin1981,Veilleux1987}, the most widely used diagram being $\oiii/\Hb$ versus $\nii/\Ha$\footnote{Throughout this work we will use the denomination \oiii and \nii to refer to \Oiii and \Nii, respectively.}, now referred to as BPT.  According to these diagrams, galaxies are divided into star-forming (SF) and active galactic nuclei (AGN) hosts, with subdivisions as LINERs, Seyferts, and composite \citep[][hereafter K01, and \citealt{Kauffmann2003c}]{Kewley2001}.

\citet{Stasinska2008} showed that the BPT diagram is not able to identify a category of galaxies whose existence was inferred from stellar evolution theory: The retired galaxies, which have the same location as LINERs in the BPT diagram. Retired galaxies are objects where star formation has stopped long ago. If they contain gas, they show emission lines  which are the result
 of photoionization by hot, low-mass evolved stars (HOLMES). \citet[][hereafter CF10 and CF11]{CidFernandes2010, CidFernandes2011} proposed a new diagram,
the WHAN diagram (the \Ha equivalent width, $W_{\Ha}$, versus $\nii/\Ha$), which is able to discriminate between weak AGN and retired galaxies.  Since then, many studies have found evidence for the existence of retired galaxies \citep[e.g.][]{Sarzi2010, Singh2013, Belfiore2015MNRAS449867B, Penny2015arXiv150806186P}, although numerous studies still ignore them. As shown by \citet{Stasinska2015MNRAS}, disregarding this category leads to an erroneous census of galaxy types.

The main aim of this paper is to show that a large fraction of BPT-LINERs, which are often considered as galaxies with a scaled-down nuclear activity, is actually composed of objects whose emission lines are naturally explained by their old stellar populations and do not require any special `activity', be it due to an accreting black hole or to shocks or to whatever was suggested in the first place to explain LINERs \citep{Heckman1980}.

An independent point raised by several authors is that diagnostics based on emission-line diagrams, when applied to spectroscopic data, as for example those from the Sloan Digital Sky Survey \citep[SDSS;][]{York2000}, lead to a classification that is aperture-dependent and may not reflect the real nature of the galaxies \citep{Gomez2003ApJ584210G, Brinchmann2013MNRAS4322112B, Papaderos2013, Stasinska2015MNRAS, Gomes2015arXiv151101300G}.

These two issues, the real nature of the BPT-LINER galaxies and the aperture-dependence of the spectral classification methods, lead us to consider a new quantity, not related to emission lines: the infrared colour. Studies dealing with the mid-infrared photometry of galaxies already noticed the bimodality in infrared colours, separating objects containing warm dust (attributed to star formation) from those devoid of it \citep[e.g.][]{daCunha2008, daCunha2010MNRAS403, Alatalo2014ApJ794L}. Here we take advantage of this fact by considering data from the Wide-field Sky Survey Explorer \citep[WISE;][]{Wright2010}.  WISE photometry has another advantage: the measurements refer to the entire galaxies and not only to the parts sampled by a spectroscopic fibre.
 
This paper is organised as follows: Section \ref{sec:data} describes the data and proposes a mid-infrared criterion to separate SF galaxies from those not forming stars. 
 It also introduces the WHAN and BPT-based spectral classes used throughout this work.
Section \ref{sec:whaw} presents a new diagram based on $\mathrm{W_{H\alpha}}$ and the WISE colour \ww. Section~\ref{sec:discussion} discusses the aperture effects and some properties of the LINER-like galaxies. Section~\ref{sec:summary} summarises our results.

\section{Preliminaries}\label{sec:data}
 
\subsection{The master sample}\label{sec:sample}

This study draws data from the $7^{\mathrm{th}}$ Data Release of the SDSS and from the WISE catalog. The SDSS gathered 3800--9200\,\AA\ spectra through fibres of 3\,arcsec in diameter for over 900\,000 galaxies \citep{Abazajian2009}. We use the emission line fluxes obtained after subtracting the stellar continuum derived with {\sc starlight} \citep{CidFernandes2005}.

WISE mapped the entire sky in four bands: $W1$, $W2$, $W3$ and $W4$, with effective wavelengths of 3.4, 4.6, 12 and 22\,$\mu$m, respectively. Most SDSS galaxies have a counterpart in WISE. $W3$ contains strong PAH emission bands, indicative of the presence of warm dust usually associated with star formation, while $W1$ and $W2$ are dominated by stellar emission \citep[e.g.][]{Meidt2012ApJ74417M, Cluver2014ApJ78290C}. Warm dust is also detected in the $W4$ band but with poorer S/N ratio. The $W1$ band is mainly related to stellar emission but contains a strong PAH band which complicates its interpretation in terms of star formation (see \citealt{Jarrett2011} for more details about the WISE bands and \citealt{Draine2007} and \citealt{daCunha2008} for PAH bands). 

From the SDSS--WISE dataset, we consider all galaxies in the SDSS Main Galaxy Sample \citep{Strauss2002}.  We further require S/N $ > $ 3 in both $W2$ and $W3$, and impose a redshift limit of $z < 0.2$ to circumvent the need for K-corrections of the WISE data.  This characterises our master sample, which contains 447\,872 galaxies.


\subsection{WISE separation between SF and lineless galaxies}\label{sec:divisor}

\begin{figure}
   \centering
   \includegraphics[width=0.95\columnwidth, trim=20 20 0 60]{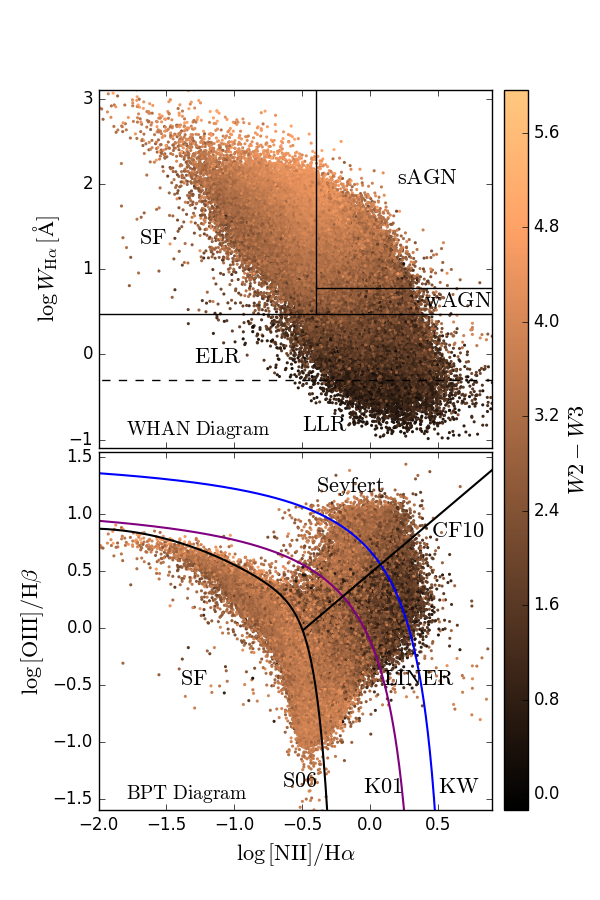}
   \caption{Top: WHAN diagram; the lines delimit the spectral classes defined in \citet{CidFernandes2011}. Bottom: BPT diagram; the black curve is the SF/AGN separator from \citet{Stasinska2006}; the purple line is the `pure AGN' classifier of \citet{Kewley2001}; the blue line is the Kozie{\l}-Wierbowska et al. (in preparation) proposed divisor; and the straight black line is the \citet{CidFernandes2010} Seyfert/LINER divisor. In both diagrams points are colour coded according to $W2-W3$.}
   \label{fig:bpt-whan}
\end{figure}

 \begin{figure*} 
   \centering
   \includegraphics[width=0.94\textwidth, trim=20 20 40 40]{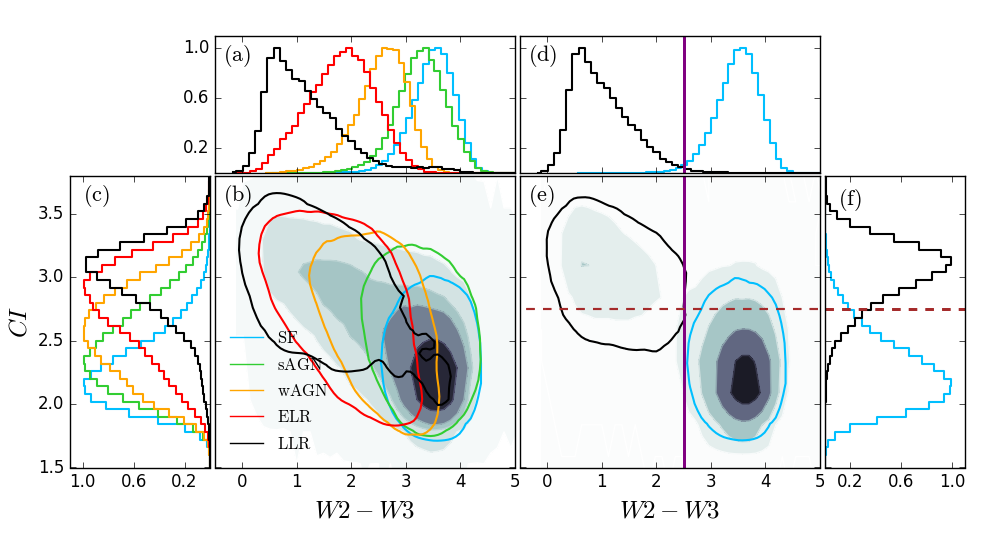}
   \caption{Panel (a): Histograms of the WHAN spectral classes for the \ww colour. All the histograms in this figure are normalized by the peak of their corresponding distribution. Panel (b): The WHAN classes as distributed in the $\ww$ versus $CI$ plane. The contours represent the 95th percentile of each WHAN class, as indicated by the legend. The shadowed regions contain 90, 75, 50 and 25 percent (from outside to inside, respectively) of all selected galaxies in the diagram. Panel (c): Same as (a) for the $CI$ parameter. Panels (d), (e) and (f): Same as (a), (b) and (c), respectively, but only for SF and elliptical LLR galaxies, with the purple straight solid line representing our best divisor between presence and absence of star formation ($\ww = 2.5$), and the straight dashed brown line marking the best divisor between SF and LLR galaxies with respect to the concentration index ($CI=2.75$).}
   \label{fig:divisor}
\end{figure*}

Given its sensitivity to the presence of warm dust powered by regions of star formation,  the \ww colour may be used to distinguish galaxies with and without star formation.  \citet{Cluver2014ApJ78290C}, for instance, in their discussion of the location of different kinds of sources in the \ww versus $W1 - W2$ colour-colour diagram, note that systems dominated by star formation have $\ww > 1.5$ Vega magnitudes (the same system used throughout this work). In this section we revisit this issue by examining how \ww relates to galaxy types as defined by their optical emission lines, with the specific goal of deriving a \ww value which best separates systems with star formation from those without it.

We base our analysis on the WHAN classification scheme of \citetalias{CidFernandes2011}, in which galaxies
 with $W_{\Ha} < 3$ \AA\ are identified as retired galaxies, i.e., systems which have stopped forming stars long ago and whose ionising photon budget is dominated by HOLMES.
 Galaxies satysfying this criterion represent about thirty percent our master sample.
In some of them emission lines are not even convincingly detected. It is thus useful to subdivide this category according to the presence or absence of emission lines. 
We call `line-less retired' galaxies (LLR) those systems where \Ha is weaker than 0.5 \AA\ in equivalent width and 
`emission-line retired' galaxies (ELR) those with \wha between 0.5 and 3 \AA\ (the limits being the same as defined in \citetalias{CidFernandes2011}).\footnote{N.B.: Our ELR and LLR classes were called `retired' and `passive' in \citetalias{CidFernandes2011} and earlier works of our group. The term `retired' was coined by  \citet{Stasinska2008} to identify emission-line galaxies that had stopped forming stars and  whose line ratios could be explained by HOLMES.  Later, \citetalias{CidFernandes2011} and \citet{Stasinska2015MNRAS}, when considering galaxies without emission lines,  used the word `passive'. This choice may have been confusing since, strictly speaking, both classes correspond to objects that are `retired' in the sense of star formation  and passively evolving.}

For the emission line classification we impose S/N $ > 3$ for H$\alpha$ and \nii for all but LLR galaxies. The resulting 403\,015 objects are represented in the WHAN diagram in Fig.~\ref{fig:bpt-whan}, where points are colour-coded by the value of \ww. Also shown are the lines that delimit the other spectral classes: SF for star-forming, sAGN for strong AGN, wAGN for weak AGN, and ELR.
 
  Note that the diagram includes a small number (5713) of galaxies in the LLR region. These sources satisfy the S/N $> 3$ cut, but their lines are so weak that they fall into our lineless bin. More importantly, note that the objects with the smallest values of \ww lie in the region of retired galaxies (with or without emission lines), as expected.

We now turn our attention to SF and LLR galaxies, the two extreme classes in our classification scheme. We emphasise that, because of the S/N cut, not all LLR galaxies appear in the WHAN diagram in Fig.\ \ref{fig:bpt-whan}. In fact, as the name suggests, most LLRs show no lines at all. To obtain a more representative sample of this class we consider all LLR galaxies from the master sample. To ensure that the lack of emission in H$\alpha$ is not due to spurious effects, such as bad pixels or sky lines, we select only those galaxies having at least 75\,percent of good pixels within $\pm 1\,\sigma$ of the centre of the gaussian used to fit the line. This gives us 32\,897 LLR galaxies\footnote{This number is significantly lower than the number of LLR galaxies the Main Galaxy Sample of the SDSS-DR7 because of the $\mathrm{S/N}(W3) > 3$ criterion.}.

Fig.~\ref{fig:divisor}a shows the distribution of \ww for the different WHAN categories. The plot reveals that \ww increases steadily along a LLR--ELR--wAGN--sAGN--SF sequence. This same ordering is observed for several other properties (see Figure 10 in \citetalias{CidFernandes2011}), including the concentration index ($CI$, defined as the ratio between the 90 and 50 percent $r$-band Petrosian radii), as seen in Fig.~\ref{fig:divisor}c. The combined distribution of points in the \ww versus $CI$ plane (panel b) reflects the well-known prevalence of SF spectral classes among late-type galaxies and of LLR among early-type galaxies.

We will use the two classes occupying opposite ends of the \ww distribution, LLR and SF (in black and blue, respectively), to determine the best \ww separator between galaxies with and without star formation. Before that, 
however, we need to understand the nature of the low amplitude bump in the distribution of LLR systems at \ww  between 3 and 4. As further discussed in Section~\ref{sec:aperture}, this apparent contradiction between the emission lines which indicate galaxy retirement and \ww which indicate star formation can be easily explained: The sources occupying the bump are galaxies with star-forming disks for which only the old `retired' bulge is covered by the SDSS fibre.

\begin{figure*} 
    \centering
    \includegraphics[width=1\textwidth]{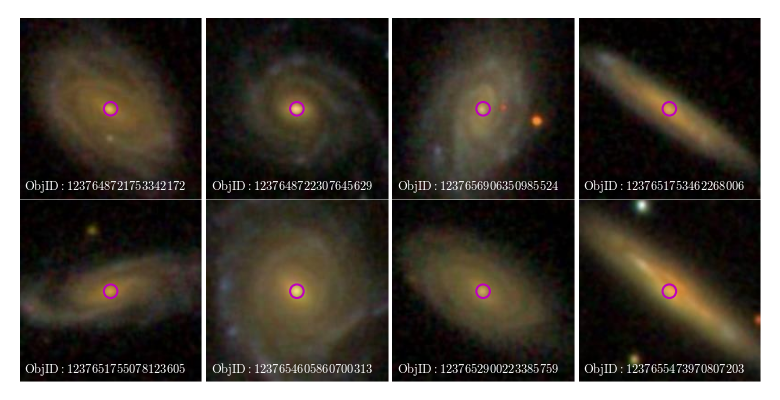}
    \caption{ Examples of lineless retired (LLR) galaxies whose $\ww$ colours are
 typical of SF galaxies (these objects are in the tail of the black histogram shown in Fig.~\ref{fig:divisor}a).  The purple circles indicate the positions of the  3 arcsec SDSS fibres. These galaxies have retired bulges (hence the LLR classification) but star-forming disks (hence the $\ww$ colour).}
    \label{fig:bulge_retireds}
\end{figure*}

A first clue in this sense is given by distribution of these systems in the $CI$ versus \ww plane in Fig.~\ref{fig:divisor}b. While most LLRs have $CI$ values indicative of early-type morphology, those with SF-like \ww colours are clearly late-type systems. This is further confirmed in Fig.~\ref{fig:bulge_retireds}, where SDSS images of a few randomly selected examples are shown. Clearly, the spectral classification of galaxies based on the 3 arcsec fibre spectra refers only to their old, retired bulge, missing the star-forming disk responsible for the WISE emission.

To clean our LLR sample of this intruding population we resort to data from the Galaxy Zoo 1 catalog \citep[]{Lintott2011MNRAS}. We keep in our LLR sample only those objects for which the probability of being elliptical, according to the catalog's criteria, is greater than 0.5. The 23\,142 elliptical galaxies in this list (70 percent of our original LLR sample) compose our `bona fide' LLR sample. The distribution of \ww and $CI$ values for the cleaned sample is shown in panels d, e and f of Fig. \ref{fig:divisor}. As expected, the bump at large in \ww for LLRs (black lines) is gone.

We then follow \citet{Strateva2001} and define two parameters: the completeness, $\mathcal{C}$, and the reliability, $\mathcal{R}$. The optimal divisor is then obtained by maximising the product $\mathcal{P} = \mathcal{C_{\mathrm{SF}}}\mathcal{R_{\mathrm{SF}}} \mathcal{C_{\mathrm{LLR}}}\mathcal{R_{\mathrm{LLR}}}$, as done before in \citet{Strateva2001}, \citet{Mateus2006}, and \citetalias{CidFernandes2010} in other contexts. This gives  $CI = 2.75$ as the value which best separates LLR from SF galaxies (\citealt{Strateva2001} and \citealt{Mateus2006} had found $CI = 2.63$ and $2.62$, respectively). As seen in Figs.~\ref{fig:divisor}e and f (where this limit is marked by a dashed horizontal line), morphology, \ww and optical emission line properties are all inter-related.

Applying the same procedure to WISE colour yields $\ww = 2.5$ mag as the best separator between galaxies with and without star formation. This value (marked by the vertical line in Figs.~\ref{fig:divisor}d and e) is 1 mag larger than the one suggested by \citet{Cluver2014ApJ78290C} on the basis of a WISE colour-colour diagram. As seen in Fig.~\ref{fig:divisor}, this relatively large difference has practically no impact  insofar as LLR galaxies are concerned, since the peak of their \ww distribution falls well below both proposed thresholds. Tagging $\ww > 1.5$ galaxies as hosting star-formation would nevertheless fail to recognise most ELRs as non-star-forming systems, while our proposed $\ww = 2.5$ divisor does a much better job in this sense. In any case, because it is explicitly tied to a classification scheme based on emission lines, our empirically derived $\ww = 2.5$ threshold is more relevant in the context of this work.

\subsection{The BPT-LINER samples}
\label{sec:liners}

Let us now momentarily forget about WHAN-based classes and define a `LINER' sample in a way similar to what is commonly done in the literature using the BPT diagram. Diagrams involving \sii or \oi lines (as proposed by \citealt{Kewley2006}) are not used here since this would reduce the size of the LINER sample without changing its properties (\citetalias{CidFernandes2010}). Imposing a standard $\mathrm{S/N} > 3$ to all four lines (\oiii, \Hb, \nii, and \Ha) leaves us with a BPT sample containing 274\,444 galaxies, shown in the BPT diagram of Fig.~\ref{fig:bpt-whan} (bottom panel), also coloured by \ww. The region with the smallest \ww values corresponds to the LINER region, especially its high \nii/\Ha end. The values of \ww are much higher along the SF wing, with the largest ones reached in the region of lowest \oiii/\Hb, populated by the most massive and metal-rich SF galaxies -- precisely the ones expected to have the strongest warm-dust emission.
 
Now we define the samples to be considered in the next sections, which contain only LINER-like galaxies from
 the BPT diagram. To do this, we use as main delimiters the `pure SF'/AGN division line of
 \citet[][hereafter S06]{Stasinska2006} and the Seyfert/LINER division line of \citetalias{CidFernandes2010}
 (both shown in bottom panel of Fig.~\ref{fig:bpt-whan}). This leaves us with a BPT-LINER sample of 129\,606
 galaxies. Finally, we define three subsamples which differ by the position of their left boundary. According
 to photoionisation models (\citetalias{Stasinska2006}, \citealt{Kewley2013ApJ774100K}) the farther the objects
 lie from the S06 line, the smaller the contribution of star formation to the hydrogen lines. This contribution
 is about 30 percent for the K01 line (usually wrongly considered as the delimiter of `pure' AGN). Our first subsample
 is limited by the S06 line, the second one is limited by the K01 line and the third one by a line displaced to
 the right of the K01 line proposed by Kozie{\l}-Wierbowska et al. (in preparation; hereafter KW). All these
 delimiters are plotted in Fig.~\ref{fig:bpt-whan}. The corresponding subsamples are referred to as S06-LINERs,
 K01-LINERs and KW-LINERs, the latter being expected to be very little affected by star formation. The number
 of objects in each subsample is 129\,606 for S06, 22\,195 for K01 and 3\,085 for KW.


\section{The WHAW diagram}
\label{sec:whaw}

\begin{figure*}
    \centering
    \includegraphics[width=0.92\textwidth]{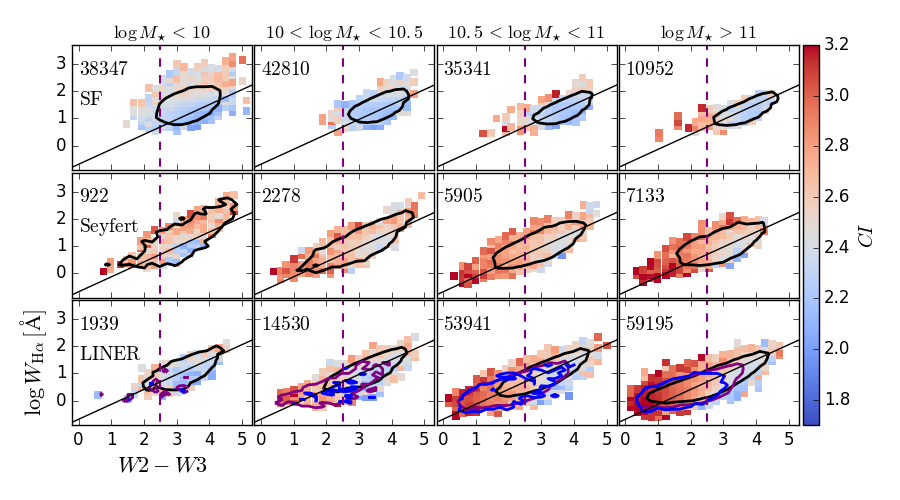}
    \caption{The WHAW diagram in bins of stellar mass (identified on top, in $M_\odot$),
             coloured by the concentration index, whose scale is indicated by the colour bar on the right.
             Upper panels: BPT S06-SF galaxies. Middle panels: BPT S06-Seyferts. Lower panels:
             BPT S06-LINERs. The straight line represents the
             $\log W_{\Ha} = 0.55 \times (\ww) - 0.69$ fit to all BPT galaxies in the
             WHAW space. The black contours in each panel mark the 90th percentiles.
             The dashed purple line represents the $\ww = 2.5$~mag divisor. The purple and blue contours in bottom panels represent the 90th percentile of K01 and KW-LINERs, respectively.}
    \label{fig:whaw}
\end{figure*}

To investigate the segregation of infrared colours in the WHAN and BPT diagrams, we use the \wha versus \ww plane
(hereafter, the WHAW diagram). As found by previous works, the \ww colour is related to the specific star formation rate (sSFR) of galaxies, both for SF galaxies and AGN hosts \citep[e.g.][]{Donoso2012ApJ74880D}. This arises from the fact that the 12 $\mu$m  emission ($W3$ band) is correlated with the star formation rate (SFR), as shown by \citet{Cluver2014ApJ78290C}, for instance, and the 4.6 $\mu$m ($W2$ band) is expected to strongly correlate with stellar mass. \wha has a different meaning for different types of objects. For pure SF galaxies it measures the specific star-formation rate; for AGN hosts it includes the ionisation by the active nucleus; and for retired galaxies it measures the ionisation by HOLMES. Therefore, at least for SF galaxies, the WHAW diagram must show a clear correlation, as both quantities involved are related to the sSFR. Besides, the WHAW diagram can tell us about aperture effects acting on the spectral classification, since \wha is based on SDSS data, so it is aperture-dependent, whilst the WISE \ww colour is not.

In Fig.~\ref{fig:whaw} we show the WHAW diagram for different classes of galaxies selected by using the S06 line in the BPT diagram: S06-SF galaxies (top panels); S06-Seyferts (middle); and S06-LINERs (bottom). Galaxies are also split in bins of stellar mass. The coloured points show the dependence of \ww on the concentration index. We see that objects with higher $CI$ values (i.e. those with early-type morphologies) tend to have lower \ww values  for all the classes considered here. 

The location of the points with respect to the best-fit relation for the whole BPT sample (black line) shows that the relation between \wha and \ww for SF galaxies depends on the galaxy mass. This is expected, since \ww increases with the galaxy dust content, which is known to increase with galaxy mass \citep{Donoso2012ApJ74880D, Clemens2013MNRAS433695C, Zahid2014ApJ79275Z}. Thus, galaxies with lower dust content tend to have a lower \ww\ for the same sSFR (or \wha). Low mass SF galaxies also show a broader relation compared to high mass objects, which can be explained in terms of their low-metallicity regime, since the SFRs derived from the 12 $\mu$m emission tend to be underestimated for metal-poor objects \citep[e.g.][]{Lee2013}.

For Seyfert galaxies, the relation is almost independent of stellar mass, with $\sim 80$ percent of the galaxies having mid IR colours indicative of ongoing star formation ($\ww > 2.5$)
 This points towards a connection between the processes ruling star formation and the efficiency of AGN activity, possibly dependent on the amount of cold gas available for both phenomena. For instance, \citet{Rosario2013ApJ77894R} show that, for the case of Seyferts, a tight correlation can be seen between the star formation rate, as estimated from the 12$\mu$m emission, and the nuclear \oiii luminosity, the latter being a measure of nuclear luminosity \citep[e.g.][]{Kauffmann2009}. However, selection effects primarily due to the limited aperture used by the SDSS to derive emission lines could play a crucial role in driving such correlation.

In the case of LINERs, the most conspicuous feature in the WHAW diagram is that they cover a wide range of \ww values and occupy both sides of the line separating SF galaxies from galaxies without current star formation. The range of \ww values is larger for large masses. Low-mass LINERs occupy the same region as  SF galaxies. In the last row of panels in Fig.~\ref{fig:whaw} we mark the contours corresponding to the S06- (black),  K01- (purple) and KW-LINER (blue) samples. We see that the \ww colour distribution for LINERs is both dependent on the stellar mass of their hosts and on the BPT criterion adopted to select them. The high mass bin shows that KW-LINERs, and to a smaller extent the K01-LINERs, are mostly found at \ww values indicating  no current star formation, implying another origin for their emission lines. In the next section we discuss these trends.

\section{Discussion}
\label{sec:discussion}

Armed with the optimal \ww separator of star-forming and retired systems obtained in Section~\ref{sec:divisor}, the BPT-based LINER samples defined in  Section \ref{sec:liners}, and the WHAW diagram introduced in Section \ref{sec:whaw}, this section revisits the nature of conventionally defined LINERs and the aperture effects involved.

\begin{figure*}
    \centering
    \includegraphics[width=0.95\textwidth]{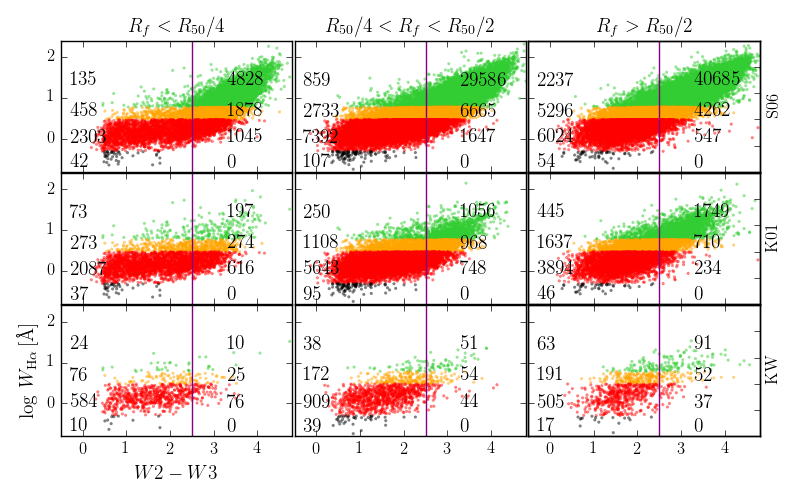}
    \caption{\ww versus $W_{\Ha}$ (the WHAW diagram) for our three BPT-LINER
      subsamples. Top, middle and bottom panels show S06-, K01- and
      KW-LINERs, respectively.  Panels from left to right correspond
      to different bins in SDSS fibre radii ($R_f$) in units of the
      corresponding $r$-band Petrosian radius ($R_{50}$). The colours
      follow the same scheme as in Fig. \ref{fig:divisor}: green points
      are sAGN, orange are wAGN, red are ELR, and black are LLR
      galaxies. The vertical line marks our optimal separation between
      SF and LLR galaxies. The numbers inside the panels indicate the
      number of objects in the different WHAN classes on each side of
      the vertical line.}
    \label{fig:ewhaise}
\end{figure*}

Fig.~\ref{fig:ewhaise} shows our three BPT-based LINER sub-samples (S06, K01 and KW) in the WHAW diagram.  Each subsample is divided according to the value of $R_f$, the galactic radius corresponding to the 1.5 arcsec radius of the SDSS fibre. Three bins are considered: (i) $R_f < R_{50}/4$, (ii) $R_{50}/4 < R_f < R_{50}/2$, and (iii) $R_f > R_{50}/2$, where $R_{50}$ is the Petrosian photometric radius within which  50 per cent of the $r$-band light of a galaxy originate.
 In this figure we follow the colour code of Fig.~\ref{fig:divisor} to identify the WHAN spectral classes. To the right of our \ww division line are the galaxies with ongoing star formation (they have warm dust and may or may not host an AGN), while to the left are objects without ongoing star formation. In all panels we indicate the number of objects in each WHAN class to the left and to the right of the $\ww = 2.5$ line. The \wha axis indicates if the gas ionisation requires source besides HOLMES. 

\subsection{Impact of aperture effects in the BPT and WHAN classifications}
\label{sec:aperture}

Let us first explore the fact that the WISE photometry does not suffer from aperture effects to study how the limited size of the SDSS fibre affects the spectral classification.

From Fig.~\ref{fig:ewhaise} we see that a fraction of ELR galaxies have high values of \ww, indicating the presence of warm dust. The incidence of these conflicting signs ($W_{\Ha} < 3$ \AA\ pointing to a retirement regime but \ww values signalling ongoing star formation) is clearly related to fibre coverage. Let us look at the central row of Fig.~\ref{fig:ewhaise}, which plots galaxies above the K01 line in bins of SDSS fibre coverage. From the numbers listed in the panels, the proportion of ELR galaxies with $\ww > 2.5$ decreases from 23 to 12 and 7 percent for fibres covering $< 1/4$, 1/4 to 1/2, and $> 1/2$ of $R_{50}$, respectively. This pattern holds true for the other LINER subsamples, although with different fractions.

This all happens because the WHAN classification relies on fibre spectra, which, for small values of $R_f/R_{50}$, cover only the old bulge, while external regions rich in star-formation are not seen. It is for this same reason that in Section~\ref{sec:divisor} it was found that galaxies being predominantly of late-type can be tagged as retired when using the WHAN diagram \citep[as recently observed by][]{Gomes2015arXiv151100744G}. 

For the redshift range that we consider ($z < 0.2$), the small fibre coverage problem will affect all galaxies. This is specially true for massive face-on disc galaxies, in which the fibre misses the spiral arms and their young stellar populations. In fact, even for the highest distances in the sample, the coverage of the galaxies will be only of a few kpc, so the BPT  and WHAN spectral classification are necessarily aperture-dependent. As $R_f$ increases, of course, the SDSS spectra refer to larger portions of the galaxies, and the signatures of the AGN and old bulge become weaker. For early-type galaxies the aperture effect will be less dramatic, given that they have a more spatially uniform distribution of stellar populations \citep{Gonzalez-Delgado2015AA581A103G}.

\subsection{The nature of LINER-like galaxies in the WHAW diagram}
\label{sec:ewhaise}

Let us now inspect Fig.~\ref{fig:ewhaise} along its vertical direction and examine how the different sub-samples of BPT-LINERs populate the WHAW space. Qualitatively, the results are the same whatever the range of apertures chosen. In the following, we give numbers for the first column of plots only.

The figure shows that most S06-LINERs (top panels) exhibit \ww\ values indicative of ongoing star formation. The occurrence of star-formation in these galaxies is not surprising given that the S06 line is built to mark the upper boundary of \textit{pure} SF systems in the BPT plane. `Composite' objects, where both star-formation and nuclear activity or other line excitation processes operate, are thus expected to be present in such a  sample. Nearly 80 percent of our S06-LINERs have $\ww > 2.5$. The \ww\ values of many K01- (and even KW-LINERs)  also point to star formation in spite of the fact that these objects are commonly considered as  `pure AGNs'. This is in agreement with the photoionization models of \citetalias{Stasinska2006} which show that as much as 70 percent of the \Ha emission in objects on the K01 line can  be due to massive stars. The incidence of SF-like \ww colours among these presumably purer-AGN subsamples is nonetheless smaller than for S06-LINERs. In fact, the distribution of \ww changes systematically along the S06-K01-KW sequence.  Among S06-LINERs, 27 percent have $\ww < 2.5$, while this fraction increases to 69 and 86 percent for K01- and KW-LINERs, respectively. In other words, the further up the right wing in the BPT diagram, the smaller the fraction of galaxies with ongoing star-formation. 

At first sight this result fits nicely with the idea that galaxies should have emission lines increasingly dominated by nuclear activity as they move away from the star-forming sequence in the BPT plane \citep[e.g.][]{Kauffmann2009}. However, this widely-held opinion was forged prior to the awareness of the demographic relevance of retired galaxies (\citetalias{CidFernandes2011}, \citealt{Stasinska2015MNRAS}), whose emission lines are powered not by accretion onto a supermassive black hole but by HOLMES. 

It is therefore useful to evaluate how the importance of HOLMES varies along the S06-K01-KW sequence of BPT-LINERs. This is where the y-axis of our WHAW diagram comes into play. The fraction of galaxies in the (empirically and theoretically motivated)  $\wha < 3$ \AA\ and $\ww < 2.5$ HOLMES-dominated regime increases from 69 percent in the case of S06-LINERs to as much as 78 and 89 percent for K01 and KW-LINERs, respectively. (These fractions are only slightly affected by the aperture effects discussed in the previous section.) In other words, {\it the further up the right wing in the BPT diagram, the higher the fraction of retired galaxies among BPT-LINERs.}

These numbers reveal that, as far as BPT-LINER systems are concerned, galaxies traditionally taken to represent `pure' AGN (i.e., K01- and KW-LINERs) are, statistically speaking, more likely to represent systems whose emission-line properties can be fully understood in terms of old stellar population properties.

Let us be clear that we are not saying that objects with low nuclear activity do not exist. Such objects belong to our weak and  strong AGN classes (orange and green points in Fig.~\ref{fig:ewhaise}). What we argue is that many BPT-LINERs are in fact ELR galaxies, as first shown by the WHAN diagram and now confirmed by the WHAW.

\subsection{The negligible contribution of AGNs to the mid-IR emission of LINERs}

Our whole analysis neglects the potential contribution of AGNs to the WISE fluxes. To some extent, the result that the \ww colours of all AGN considered in this work lie between those of SF and LLR galaxies (see Fig.~\ref{fig:divisor}a), neither of which is suspected to host AGN,  vindicates this assumption. It is nevertheless worth addressing this issue in light of previous work which indicates that AGN emission in the mid-IR can be significant in some cases   \citep{Mateos2012MNRAS4263271M,Mateos2013MNRAS434941M,Yan2013,Caccianiga2015}.

\begin{figure}
    \centering
    \includegraphics[width=0.92\columnwidth, trim=20 30 20 20]{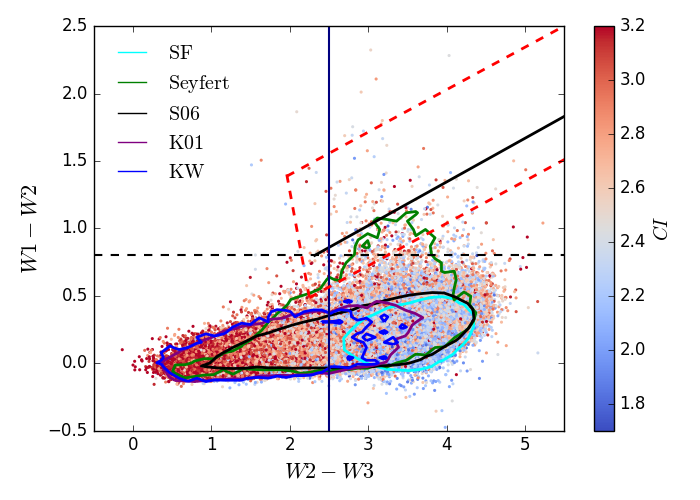}
    \caption{ $W1-W2$ versus $W2-W3$ colour-colour diagram for our BPT-LINER sample. Points are coloured as a function of concentration index $CI$. The contours indicate the 90th percentiles for each LINER sub-sample selected by using the S06, K01, and KW lines (black, purple, and blue lines, respectively). The 90th-percentile contours in cyan and green mark the loci for SF galaxies and Seyfert hosts selected with the S06 line in the BPT diagram (note that we do not plot the points for those galaxies). The vertical solid line indicates our best divisor between objects with ongoing star formation and retired galaxies. The dashed horizontal line is the mid-IR criterion to select AGNs proposed by \citet{Stern2012}. The black solid line marks the location of a simple power-law spectrum with varying slope as proposed by \citet{Caccianiga2015}, and the dashed red lines delineate the `AGN wedge' as defined by \citet{Mateos2012MNRAS4263271M}.}
    \label{fig:wise_liners}
\end{figure}

To evaluate the potential contribution of AGN to the mid-IR emission of our LINERs, Fig.~\ref{fig:wise_liners} presents a $W1-W2$ versus $W2-W3$ colour-colour diagram. The points are coloured as a function of the concentration index. In this figure, we also show the optimal divisor line obtained in this work, $W2-W3 = 2.5$, the mid-IR criterion to select AGNs proposed by \citet{Stern2012},  $W1-W2 > 0.8$,  and the locus expected for a power-law spectrum of varying slope, above which the infrared emission could be explained purely by nuclear activity \citep{Caccianiga2015}. Dashed red lines mark the `AGN wedge' defined by \citet{Mateos2012MNRAS4263271M}.

As it is clear from the plot, only a few objects in our sample lie in the AGN region. For the overwhelming majority of the BPT-LINERs, the host galaxy dominates the mid-IR emission, explaining their loci on the $W1-W2$ versus $W2-W3$ diagram.

The black, purple, and blue lines  in Fig.~\ref{fig:wise_liners} mark the 90 percent contours of S06, K01, and KW-LINERs, respectively.  A gradual shift towards smaller \ww and slightly smaller $W1-W2$ along the S06-K01-KW sequence is observed. If AGN emission contributed significantly to the mid-IR, this sequence towards purer-AGN should move sources towards the top-right of this diagram, which is not what is seen. Seyfert galaxies (green contours) do stretch towards the AGN region, indicating some contribution of the AGN to the mid-IR emission.
For  BPT-LINERs, the focus of this paper, however, we conclude that this contribution is negligible.

\section{Summary}
\label{sec:summary}

In this work we have selected a sample of galaxies from the SDSS and WISE surveys and studied their mid-infrared colour \ww, a powerful indicator of the presence or absence of ongoing star formation. We find that $\ww = 2.5$ optimally separates SF from retired galaxies.

We focused our attention on galaxies commonly tagged as LINERs in the BPT plane and constructed a new diagram, the WHAW diagram, plotting \wha versus \ww. We found that this diagram confirms that most galaxies classified as retired by the WHAN diagram (i.e. those with $\wha < 3$~\AA, which includes many BPT-defined LINERs) indeed do not present any sign of ongoing star formation in the WISE data. \citet{Stasinska2008} and \citetalias{CidFernandes2011} showed that the optical emission lines of these systems are explained by their populations of HOLMES and do not require the presence of an AGN. 

In addition, we warn against interpreting the entire right wing in the BPT diagram as a SF--AGN mixing line. By studying subsamples of BPT-LINERs farther away from the SF sequence, we conclude that the tip of the BPT-LINER wing is dominated by emission-line retired galaxies powered by HOLMES, and not by AGNs.

The WISE data also allowed us to tackle aperture effects, which may bias galaxy classifications based on fibre spectroscopy. For galaxies with a small covering fraction, a non-negligible fraction of retired galaxies have a \ww colour indicative of star formation. In these cases, the WHAN diagram relates to a `retired' bulge, while the \ww colour refers to the entire galaxy and indicates the presence of SF regions in the galactic disk.
In conclusion, while WHAN-retired SDSS galaxies may be counterfeiters of retired bulges in SF galaxies, WISE-retired galaxies are truly retired.

\section*{Acknowledgements}

We thank the anonymous referee for the useful suggestions to improve this paper. We acknowledge the support from the Brazilian agencies FAPESC, CAPES, CNPq, and the CAPES CsF--PVE project 88881.068116/2014-01. The {\sc starlight} project is supported by Brazilian agencies CNPq, CAPES, by the France--Brazil CAPES--COFECUB programme, and by Observatory de Paris. 

This publication makes use of data products from the Wide-field Infrared Survey Explorer, which is a joint project of the University of California, Los Angeles, and the Jet Propulsion Laboratory/California Institute of Technology, funded by the National Aeronautics and Space Administration.

Funding for the SDSS and SDSS-II has been provided by the Alfred P. Sloan Foundation, the Participating Institutions, the National Science Foundation, the U.S. Department of Energy, the National Aeronautics and Space Administration, the Japanese Monbukagakusho, the Max Planck Society, and the Higher Education Funding Council for England. The SDSS Web Site is http://www.sdss.org/.

The SDSS is managed by the Astrophysical Research Consortium for the Participating Institutions. The Participating Institutions are the American Museum of Natural History, Astrophysical Institute Potsdam, University of Basel, University of Cambridge, Case Western Reserve University, University of Chicago, Drexel University, Fermilab, the Institute for Advanced Study, the Japan Participation Group, Johns Hopkins University, the Joint Institute for Nuclear Astrophysics, the Kavli Institute for Particle Astrophysics and Cosmology, the Korean Scientist Group, the Chinese Academy of Sciences (LAMOST), Los Alamos National Laboratory, the Max-Planck-Institute for Astronomy (MPIA), the Max-Planck-Institute for Astrophysics (MPA), New Mexico State University, Ohio State University, University of Pittsburgh, University of Portsmouth, Princeton University, the United States Naval Observatory, and the University of Washington.

\bibliographystyle{mnras}
\bibliography{library}

\bsp
\label{lastpage}

\end{document}